\documentstyle{aipproc}

\newcommand{\be}{\begin{equation}}   \newcommand{\ee}{\end{equation}}
\newcommand{\bear}{\begin{eqnarray}}
\newcommand{\eear}{\end{eqnarray}}
\newcommand{\ba}{\begin{array}}      \newcommand{\ea}{\end{array}}
\newcommand{\lae}{\begin{array}{c}\,\sim\vspace{-21pt}\\< \end{array}}

\newcommand{\up}{\mbox{$P$}}
\newcommand{\down}{\mbox{$N$}}
\newcommand{\inl}{{\scriptscriptstyle L}}
\newcommand{\inr}{{\scriptscriptstyle R}}
\begin{document}
\title{Testing Technicolor with Scalars \\ [2mm] at the First Muon 
Collider\footnote{Talk presented at the Workshop on Physics at the First 
Muon Collider, Fermilab,  Nov. 6-9, 1997.} }

\author{Bogdan A. Dobrescu}
\address{Fermi National Accelerator Laboratory\\
P.O. Box 500, Batavia, Illinois, 60510, USA} 

\maketitle

  \vspace*{-6.6cm}
\noindent
\makebox[8.2cm][l]{hep-ph/9802259} FERMILAB-CONF-98/044-T \\ [1mm]
\makebox[7cm][l]{February 5, 1998} \\ 
 \vspace*{5.6cm}

\begin{abstract}
An interesting class of models of dynamical electroweak symmetry breaking 
allows only the third generation fermions to acquire dynamical masses,
such that the masses of the first two generations should be given by coupling
to a nonstandard ``Higgs'' doublet. The scalars in this case have large 
couplings to the second generation, so that they are copiously produced
at a muon collider.
We analyze the potential for discovery of the neutral scalars in the 
$s$-channel, and we show that at resonance there will be observed in excess
of $10^5$ events per year.
\end{abstract}

\section*{Introduction}

Although the electroweak symmetry breaking and the quark and lepton spectrum
may have a common origin, it is quite possible that they are generated by
some new physics which manifests itself at low energy as distinct sectors.
For example,  the effective theory at a scale of order 1 TeV can include a 
sector that gives rise to the masses of the $W$, $Z$ and 
third generation fermions, as well as a sector responsible
for the masses of the lighter fermions.

This is the case when the electroweak symmetry is broken dynamically 
by some new strong gauge interactions, and only the third generation quarks 
and leptons couple to the fields charged under these interactions.
In the context of technicolor models, this situation is discussed in 
\cite{susytc,techniscalars}. 
The minimal mechanism for producing the masses of the first two generations 
requires a weak-doublet scalar transforming as the Higgs doublet, but having
a positive squared-mass, as in the technicolor models with a scalar 
\cite{tcscalar,carsim} or in bosonic technicolor models 
\cite{bosonic,horizontal}.

By coupling to the dynamical symmetry breaking sector, the scalar acquires a 
small vacuum expectation value (VEV).
The existence of the scalars is not troublesome as long as the theory described
here is the low energy manifestation of a TeV scale theory which includes
scalar compositness or softly broken supersymmetry.

Because the VEV of the weak-doublet scalar is smaller than the electroweak 
scale, the Yukawa couplings of the ``nonstandard-Higgs'' bosons to the first 
and second generations must be larger than in the standard model. 
This is relevant for future collider searches. In particular, since these
Yukawa couplings are proportional to the fermion mass, the $s$-channel 
production is very large at a muon collider.

Here we discuss briefly the discovery potential and resonant production of the 
``nonstandard-Higgs'' bosons at a First Muon Collider, operating 
at a center of mass energy of up to 500 GeV. 
A comprehensive study of the collider phenomenology of the 
``nonstandard-Higgs'' will be given elsewhere
\cite{collider}.

\section*{Technicolor with Scalars}

We assume that the electroweak symmetry breaking and the third generation 
fermion masses have a dynamical origin. To be specific, we consider a 
technicolor model constructed along the lines of ref.~\cite{techniscalars}.
The second and first generation fermions acquire masses by 
coupling to a scalar, $\phi$, which transforms under the gauge group like the
standard model Higgs doublet.

In addition to the standard model fermions, consider one doublet
of technifermions, \up\ and \down, and three scalars, $\omega$, $\chi$ and 
$\phi$, which transform under the $SU(4)_{\rm TC} \times SU(3)_{\rm C}
\times SU(2)_{\rm W} \times U(1)_{\rm Y}$ gauge group as:
\bear
& & \!\!\!\! \Psi_{\!\!\inr} =
\left(\!\!\ba{c}\up_{\inr} \\ \down_{\inr}\ea\!\!\right)
  \, : \; (4, 1, 2)_0 \; , \hspace{.5cm}
\up_{\inl} \, : \;  (4, 1, 1)_{+1}
\; , \hspace{.5cm}
\down_{\inl} \, : \;  (4, 1, 1)_{-1} \; ,
\nonumber \\ [4mm]
& & \!\!\!\! \omega \, : \; (4, \overline{3}, 1)_{- \frac{1}{3}}
\; , \hspace{.5cm}
\chi \, : \; (4,  1, 1)_{+1} \; , \hspace{.5cm} 
\phi \, : \; (1,  1, 2)_{+1} ~.
\label{e1}
\eear

The most general Yukawa interactions of $\omega$ and $\chi$ include only terms 
linear in the quark or lepton fields, such that there is a particular 
eigenstate basis in which  only 
the third generation couples to the technicolored fields:
\be
{\cal L}_{\rm Y}^{\omega,\chi} = C_q \overline \Psi_{\!\!\inr} q_{\inl}^3 
\omega 
+ C_t \overline t_{\inr} \up_{\inl} \omega^{\dagger} +
C_b \overline b_{\inr} \down_{\inl} \omega^{\dagger} +
C_l \overline \Psi_{\!\!\inr} l_{\inl}^3 \chi +
C_\tau \overline \tau_{\inr} \down_{\inl} \chi^{\dagger} +
{\rm h.c.}
\ee
Using a phase redefinition on the third generation fields, $q_{\inl}^3,
t_{\inr}, b_{\inr}, l_{\inl}^3, \tau_{\inr}$,
we can choose the Yukawa coupling constants, $C_q, C_t, C_b, C_l,
C_\tau$, to be positive.

We assume that the $\omega$ and $\chi$ techniscalars are sufficiently heavy 
to be
integrated out, such that their effects in the low energy theory are given
by four-fermion operators involving two technifermions and two 
fermions of the third generation.
As in QCD, the $SU(4)_{\rm TC}$ technicolor interactions trigger the formation 
of technifermion condensates,
\be
\langle \overline{P} P \rangle \approx \langle \overline{N} N \rangle 
 \approx 2\sqrt{3}\pi f^3 ~, 
\ee
which breaks the electroweak symmetry at a scale $f$. This also results in 
masses for the $t$, $b$ and $\tau$  \cite{susytc,techniscalars}:
\bear
m_t & \approx & \frac{\sqrt{3}}{2} \frac{C_qC_t}{M_\omega^2}\pi f^3 
\; , \nonumber \\ [.2cm]
\frac{m_b}{m_t} & = & \frac{C_b}{C_t}
\; , \nonumber \\ [.2cm]
\frac{m_\tau}{m_t} & = & \frac{C_\tau C_l}{C_t C_q}
\left(\frac{M_\omega}{M_\chi}\right)^{\! 2}
~.
\label{mtop}
\eear

The $\phi$ has Yukawa couplings to the standard fermions (these are similar 
with the standard model couplings of the Higgs doublet, but with different 
coupling constants), and also to the technifermions:
\bear
{\cal L}_{\rm Y}^{\phi} & = & \lambda_{jk}^e \overline{l}_L^j e_R^k \phi + 
\lambda_{jk}^u \overline{q}_L^j u_R^k i\tau^2 \phi^\dagger + 
\lambda_{jk}^d \overline{q}_L^j d_R^k \phi 
\nonumber \\ [.2cm]
& & + 
\lambda_+\overline{\Psi}_L P_R i\tau^2 \phi^\dagger +
\lambda_-\overline{\Psi}_L N_R\phi +
{\rm h.c.}
\eear
where $i,j = 1,2,3$ are generational indices, and the $\lambda$'s are
coupling constants.
We consider the case where $\phi$ has a positive squared-mass,
$M_{\phi}^2 > 0$, so in contrast to the standard model, the $\phi$
sector does not induces a VEV by itself. However,
when the technifermions condense, the last two terms in the above 
Lagrangian give rise to tadpole terms, such that $\phi$ develops a VEV
whose magnitude is \cite{tcscalar,carsim}
\be
\frac{f^\prime}{\sqrt{2}} \approx (\lambda_+ + \lambda_-)
\frac{2\sqrt{3} \pi f^3}{M_{\phi}^2}  ~.
\label{fprimevev}
\ee
Note that we neglected a possible quartic term in the $\phi$ potential. If 
the high energy theory that accounts for the existence of $\phi$ turns out to 
produce such a quartic term (and possibly higher dimensional terms), then 
its effects can be easily included. The observed $W$ and $Z$ masses require 
\be
f^2 + f^{\prime 2} = v^2 \approx (246 \; {\rm GeV})^2 ~.
\ee
The first three terms of ${\cal L}_{\rm Y}^{\phi}$ are responsible for the 
masses of the quarks and leptons
of the first two generations, and for the CKM elements. 

The effective theory below a scale of order 1 TeV where the technicolored 
fields are integrated out includes only the standard fermions and gauge 
bosons, an iso-triplet of Nambu-Goldstone bosons, $\pi^a$, $a = 1,2,3$,
associated with the chiral symmetry breaking of the technifermions, and
the  components of $\phi$ [we assume $M_{\phi} \lae 1$ TeV, although 
eq.~(\ref{fprimevev}) does not exclude multi-TeV values].
$\phi$ decomposes into an iso-singlet $\sigma$, 
and an iso-triplet $\pi^{\prime a}$:
\be
\phi = \frac{1}{\sqrt{2}}  e^{- i \pi^{\prime a} \tau^a/f^\prime}
\left(\ba{c} 0 \\ \sigma + f^\prime \ea \right) ~.
\label{decomp}
\ee
The triplets $\pi^a$ and $\pi^{\prime a}$ mix and give rise to the 
longitudinal $W$ and $Z$, and to a triplet of physical pseudo-scalars.
However, 
the large top mass suggests that $f \approx v$, so that 
\be
f^\prime \ll f ~,
\label{ffp}
\ee
which implies that the mixing is small, and we will neglect it.
Another consequence of inequality (\ref{ffp}) is that 
\be
\lambda_+ + \lambda_- \ll \frac{1}{2\pi \sqrt{6}} \left( 
\frac{M_\phi}{f} \right)^2 ~.
\ee
In this situation, the neutral real scalars $\sigma$ and 
$\pi^{\prime 3}$, and the charged scalars, 
$\pi^{\prime\pm} = (\pi^{\prime 1} \mp i \pi^{\prime 2})/\sqrt{2}$, are 
almost degenerate,
with a mass $M_\phi$. The splittings in their masses are of order 
$(f^{\prime}/f)^2$ and $\lambda_\pm^2 (M_\phi/f)^2$.

\section*{Signals at the First Muon Collider}

If the model presented in the previous section is indeed the correct
description of physics up to a TeV scale, then the only direct discovery 
accessible at a $\mu^+\mu^-$ collider with $\sqrt{s}$ 
below the first technihadron resonance will be the existence of the
components of the $\phi$ doublet. Furthermore, for 
$M_\phi < \sqrt{s} < 2 M_\phi$,
only the neutral scalars, $\sigma$ and $\pi^{\prime 3}$, can be produced.

The couplings of $\sigma$ and $\pi^{\prime 3}$ to quarks and leptons 
are in general flavor non-diagonal.  
However, because the off-diagonal couplings are constrained by 
flavor-changing neutral current measurements, we are going to consider 
only the flavor-diagonal couplings. These are proportional with the 
corresponding fermion masses in the case of the first two generations.
The couplings to the third generation are more arbitrary, 
because they are proportional only with the 
small contributions to the fermion mass, $\delta m_f$ with $f = t, b, \tau$, 
from the $\phi$ VEV. 

Since the bulk of electroweak symmetry breaking is provided by the technicolor
sector, the couplings of the neutral scalars, $\sigma$ and $\pi^{\prime 3}$,
to $W^+W^-$ and $ZZ$ are smaller by a factor of $v/f^{\prime}$ than the 
corresponding standard model couplings of the Higgs boson.
On the contrary, the couplings of $\sigma$ and $\pi^{\prime 3}$,
to the fermions of the second and first generations are enhanced by a factor 
of $v/f^{\prime}$ compared to the standard model Higgs boson.
The couplings to the third generation are also enhanced by $v/f^{\prime}$,
but they are suppressed by $m_f/\delta m_f$.

The total decay widths of the $\sigma$ and 
$\pi^{\prime 3}$ scalars are equal, and given by
\bear
\Gamma & \approx & \frac{M_\phi}{32 \pi f^{\prime 2}} \left[
3 m_c^2 + 3 m_s^2 + m_\mu^2
+ 3 (\delta m_t)^2 \left(1 - \frac{4 m_t^2}{M_\phi^2}\right)^{\! 1/2}
\theta(M_\phi-2m_t) \right.
\nonumber \\ [2mm]
& & + \left.
3 (\delta m_b)^2 + (\delta m_\tau)^2 \right] + \Gamma(W^+W^- + ZZ) ~.
\eear
We take the VEV of $\phi$ in the range
\be
1 \; {\rm GeV} \lae f^{\prime}  \lae 10 \; {\rm GeV} ~,
\ee
where the lower bound is chosen to avoid 
Yukawa coupling constants larger than order one, and the upper bound
is chosen to satisfy condition (\ref{ffp}).
In this case, the width for scalar decay into pairs of gauge bosons,
$\Gamma(W^+W^- + ZZ)$, is at most a few percent of the width for
$\sigma,\pi^{\prime 3} \rightarrow c\bar{c}$, and we neglect it.
Generically, there is no reason to expect that 
$\delta m_f$ is larger than the corresponding second generation mass.
For simplicity we assume $(\delta m_f)^2 \ll m_c^2$, such that 
the width of the $\sigma$ and 
$\pi^{\prime 3}$ scalars is dominated only by the $c\bar{c}$ final state:
\be
\Gamma \approx \frac{3 m_c^2 M_\phi}{8 \pi f^{\prime 2}}
\approx 13.2 \; {\rm GeV} \left(\frac{3 \; {\rm GeV}}{f^{\prime}}\right)^{\! 2}
 \left(\frac{M_\phi}{500 \; {\rm GeV}}\right) ~.
\ee

Given the enhanced couplings to the second generation, the $s$-channel
production of the neutral scalars at a $\mu^+\mu^-$ collider is large.
The natural spread in the muon collider beam energy, $\sigma_{\sqrt{s}}$,
is much smaller than $\Gamma$, and can be ignored in computing the effective 
$s$-channel resonance cross section \cite{cross}:
\bear
\overline{\sigma}(\mu^+\mu^- \rightarrow \sigma,\pi^{\prime 3} \rightarrow X) 
& \approx & \frac{4\pi \Gamma^2}{(s - M_\phi^2)^2 + M_\phi^2 \Gamma^2} 
\nonumber \\ [2mm] & & \times
B(\sigma,\pi^{\prime 3} \rightarrow \mu^+\mu^-)
B(\sigma,\pi^{\prime 3} \rightarrow X) ~.
\eear
We are interested especially in the case where the final state is
$X \equiv c\bar{c}$.
The branching fractions for the decays into a pair of muons, respectively into 
 $c$-quarks, are given by 
\bear
B(\sigma,\pi^{\prime 3} \rightarrow \mu^+\mu^-) & \approx &
\frac{m_\mu^2}{3 m_c^2} \approx 0.2 \%
\nonumber \\ [2mm]
B(\sigma,\pi^{\prime 3} \rightarrow c\bar{c}) & \approx &
1 - \frac{m_s^2}{3 m_c^2} - ...
\eear
where the ellipsis stands mainly for the branching fractions into
$W^+W^-$, $ZZ$, $\mu^+\mu^-$, etc. Therefore,
\be
\overline{\sigma}(\mu^+\mu^- \rightarrow \sigma,\pi^{\prime 3} \rightarrow 
c\bar{c})
\approx \frac{8\pi \Gamma^2}{(s - M_\phi^2)^2 + M_\phi^2 \Gamma^2}
\left(\frac{m_\mu^2}{3m_c^2} \right) ~.
\ee

The main background comes from $\mu^+\mu^- \rightarrow \gamma^*, Z^*
\rightarrow c\bar{c}$, and amounts to
\be
\sigma_B(\mu^+\mu^- \rightarrow c\bar{c}) \approx 0.7 \; {\rm pb} 
\frac{(500 \; {\rm GeV})^2}{s} ~.
\ee

We first study the discovery potential of a $\mu^+\mu^-$ collider
operating at a maximum center of mass energy of 500 GeV.
The beam energy can be reduced at the expense of luminosity.
A decrease in the  beam energy by a factor of two leads to a decrease in 
luminosity by a factor of ten \cite{muonreport}. 
The number of scan points can be optimized as follows. 
Consider that the adjacent scan points are separated by an energy difference
$x\Gamma$, implying that 
\bear
\overline{\sigma}(\mu^+\mu^- \rightarrow \sigma,\pi^{\prime 3} \rightarrow 
c\bar{c}) & \ge & \frac{8\pi}{(4x^2 + 1) M_\phi^2}
\left(\frac{m_\mu^2}{3m_c^2} \right) 
\nonumber \\ [2mm]
& \approx & \frac{80 \; {\rm pb} }{4 x^2 + 1}
\left(\frac{500 \; {\rm GeV}}{M_\phi}\right)^{\! 2} ~.
\eear
To observe at a particular scan point with $\sqrt{s}$ a number of $c\bar{c}$ 
final state 
events which is $5\sigma$ over the background requires an integrated luminosity 
\bear
L(s) & \ge &
\frac{5 \sigma_B(\mu^+\mu^- \rightarrow c\bar{c})}
{r( c\bar{c}) 
[\overline{\sigma}(\mu^+\mu^- \rightarrow \sigma,\pi^{\prime 3} 
\rightarrow c\bar{c})]^2}
\nonumber \\ [2mm]
& \approx & 
\frac{5.6 \times 10^{-4} \; {\rm pb}^{-1}}{(500 \; {\rm GeV})^2} \,
\frac{(4 x^2+1)^2 M_\phi^2}{r( c\bar{c}) s} ~, 
\eear
where $r( c\bar{c})$ is the efficiency for observing the $c\bar{c}$ final state,
and is given basically by the square of the $c$-tagging efficiency.
The integrated luminosity necessary for searching the 
scalar resonance over the whole range of beam energy is
\bear
L & = & \frac{1}{x\Gamma} \int_{\sqrt{s}_{\rm min}}^{\sqrt{s}_{\rm max}} 
d(\sqrt{s}) L(s) 
\nonumber \\ [2mm]
& \approx &
\frac{5.6 \times 10^{-4} \; {\rm pb}^{-1} }{(500 \; {\rm GeV})^2} \,
\frac{(4 x^2+1)^2}{r( c\bar{c})x} \frac{M_\phi^4}{\Gamma}
\left(\frac{1}{ \sqrt{s}_{\rm min} } - \frac{1}{\sqrt{s}_{\rm max}} \right)~.
\eear
Clearly, $x$ should be chosen to minimize $L$. At the minimum, 
\be
x = \frac{1}{2\sqrt{3}} ~,
\ee 
which gives
\be
L_{\rm min} = \frac{0.14 \; {\rm fb}^{-1}}{r( c\bar{c})}
\left(\frac{f^{\prime}}{3 \; {\rm GeV}}\right)^{\! 2}
\left(\frac{M_\phi}{500 \; {\rm GeV}}\right)^{\! 3} 
\left(\frac{500 \; {\rm GeV}}{ \sqrt{s}_{\rm min} } - 
\frac{500 \; {\rm GeV}}{\sqrt{s}_{\rm max}} \right) ~.
\ee
For $\sqrt{s}_{\rm max} = 500 \; {\rm GeV}$, 
$\sqrt{s}_{\rm min} = 250 \; {\rm GeV}$, and $r( c\bar{c}) \approx 10\%$,
\be
L_{\rm min} = 1.4 \; {\rm fb}^{-1} 
\left(\frac{f^{\prime}}{3 \; {\rm GeV}}\right)^{\! 2} 
\left(\frac{M_\phi}{500 \; {\rm GeV}}\right)^{\! 3} ~.
\label{total}
\ee

The peak luminosity assumed at this workshop is 
$2\times 10^{34}$ cm$^{-2}$s$^{-1}$ at $\sqrt{s} = 500 \; {\rm GeV}$, 
which corresponds to roughly $2\times 10^{33}$ cm$^{-2}$s$^{-1}$ at
$\sqrt{s} = 250 \; {\rm GeV}$.
Note that the average luminosity is significantly lower,
$7\times 10^{32}$ cm$^{-2}$s$^{-1}$ at $\sqrt{s} = 500 \; {\rm GeV}$, 
but given that the scalar resonance is broader than the spread in 
the beam energy, there is no need for a good energy resolution 
(the upper value given in \cite{muonreport}, $\delta E/E = 0.12\%$, being 
sufficient), and therefore the peak luminosity can be used. This situation 
is in contrast with the search for a light standard model Higgs which is 
narrower than the spread in beam energy.

Eq.~(\ref{total}) shows that with a luminosity of order 
100 ${\rm fb}^{-1}$/year, the scalar resonance will be
discovered in a short period of time, even for $f^\prime$ larger than 10 GeV.

Once the resonance is found, either at the muon collider by varying the beam 
energy, or at the LHC, the beam energy can be adjusted to the 
peak (even if this requires a significant reduction in the luminosity)
and then the production cross section becomes very large:
\be
\overline{\sigma}(\mu^+\mu^- \rightarrow \sigma,\pi^{\prime 3} \rightarrow 
c\bar{c})
\approx 80 \; {\rm pb} \left(\frac{500 \; {\rm GeV}}{M_\phi}\right)^{\!\! 2} ~.
\ee
With a luminosity of $2\times 10^{34}$ cm$^{-2}$s$^{-1}$
($7\times 10^{32}$ cm$^{-2}$s$^{-1}$), and a $c$-tagging 
efficiency of $~30 \%$, there are going to be 
observed approximately $10^6$ ($5 \times 10^4$) events per year. 
This will make possible 
precision measurements of the couplings and masses of the neutral 
``nonstandard-Higgs'' bosons, which in turn will open a window towards the
TeV scale physics.

It is remarkable that the peak cross section is independent of the $\phi$
VEV. This is a consequence of the small couplings of the $\phi$ to pairs of 
gauge bosons.

\section*{Conclusions}

The large resonance cross section computed in section 3 warrants
the label ``nonstandard-Higgs factory'' for the muon collider. 
On the other hand, if the class of dynamical electroweak symmetry breaking
models discussed here is correct, and the  $\phi$ mass turns out to be 
significantly larger than 500 GeV, then
no discovery will be made at the First Muon Collider, and 
there is need for a higher-energy muon collider. As discussed repeatedly at 
this workshop, the 4 TeV muon collider would also be a 
great tool for studying the strong dynamics sector.
We emphasize that a ``nonstandard-Higgs'' with large couplings to the 
second generation might also be necessary in models 
where the strong dynamics is different than technicolor,
for example in models that incorporate the top condensation seesaw
mechanism \cite{seesaw}, or in models with discrete horizontal symmetries 
\cite{horizontal}.

\vspace*{.8cm}

{\it Acknowledgements:} I would like to thank the convenors of the 
Strong Dynamics working group, Estia Eichten and Pushpa Bhat, for 
creating a stimulating atmosphere.
I am grateful to my collaborators, Tao Han and Hong-Jian He, 
for many useful comments.

\newcommand{\np}{{\it Nucl.\ Phys.}\ {\bf B}}
\newcommand{\pr}{{\it Phys.\ Rev.}\ }
\newcommand{\prd}{{\it Phys.\ Rev.}\ {\bf D}}
\newcommand{\prp}{{\it Phys.\ Rep.}\ }
\newcommand{\prl}{{\it Phys.\ Rev.\ Lett.}\ }
\newcommand{\pl}{{\it Phys.\ Lett.}\ {\bf B}}
\newcommand{\ptp}{{\it Prog.\ Theor.\ Phys.}\ }
\newcommand{\ap}{{\it Ann.\ Phys.}\ }
\newcommand{\intl}{{\it Int.\ J.\ Mod.\ Phys.}\ {\bf A}}


\end{document}